\documentclass[a4paper]{article}
\usepackage[a4paper, total={6.4in, 9.5in}]{geometry}
\usepackage{cite}
\usepackage{algorithmic}
\usepackage{textcomp}
\usepackage{graphicx}
\usepackage{xcolor}

\usepackage[hyphenbreaks]{breakurl}
\usepackage{authblk}
\usepackage{seqsplit}
\usepackage{graphicx}

\usepackage{float}
\usepackage{soul}
\usepackage{url}
\usepackage{booktabs}
\usepackage{caption} 
\usepackage[numbers]{natbib}
\begin{document}

\title{Feature importance in mobile malware detection}
\date{}

\author[1]{Vasileios Kouliaridis}
\author[1]{Georgios Kambourakis}
\author[2]{Tao Peng}

\affil[1]{Department of Information \& Communication Systems Engineering, University of the Aegean, Greece}
\affil[2]{School of Computer Science and Educational Software, Guangzhou University,
Guangzhou, China}

\maketitle

\begin{abstract}

The topic of mobile malware detection on the Android platform has attracted significant attention over the last several years. However, while much research has been conducted toward mobile malware detection techniques, little attention has been devoted to feature selection and feature importance. That is, which app feature matters more when it comes to machine learning classification. After succinctly surveying all major, dated from 2012 to 2020, datasets used by state-of-the-art malware detection works in the literature, we analyse a critical mass of apps from the most contemporary and prevailing datasets, namely Drebin, VirusShare, and AndroZoo. Next, we rank the importance of app classification features pertaining to permissions and intents using the Information Gain algorithm for all the three above-mentioned datasets.

\end{abstract}


\section{Introduction}

Mobile devices are an integral part of our everyday life. From online social networks to mobile banking transactions, mobile devices are more or less trusted and used by billions of people. According to recent reports \cite{statcounter}, \cite{idc:online}, the Android operating system (OS) is the most prevalent mobile platform, with a market share that exceeds 74\%. On the downside, the popularity and openness of this platform makes it an alluring target for malware writers. According to a Kaspersky report, 3,5M mobile malicious installation packages for Android have been discovered in 2019 \cite{securelist}. And while this number is lesser than that of 2018, it is surely not a situation that leaves much room for complacency.

Indeed, the topic of mobile malware detection has already received a lot of attention in the literature. Current mobile malware detection approaches lean primarily towards static anomaly-based detection \cite{Kouliaridis-survey, Yan, Souri, Odusami, Narudin}, although methods based on dynamic analysis have started to proliferate \cite{Kouliaridis-survey, Odusami, Kouliaridis, Papamartz}. Generally, anomaly-based detection comprises two distinct phases; the training and the detection or testing one. It typically employs machine learning to detect malicious behavior, i.e., deviation from a model built during the training phase. Naturally, the cardinal reason behind the popularity of static analysis techniques arises from the fact that they do not require the app to be running, hence they are usually faster and straightforward to implement. In this context, a key point, which to our knowledge is not properly addressed in the hitherto literature, is the importance of each feature category, say, permissions and intents, in mobile app classification. Simply put, which group of features in general, and which features within each group in particular do contribute the most information when it comes to classification? And, is the answer to the previous question related to the employed dataset? 

To respond to the previous questions, this work first briefly surveys all the major datasets used in the context of app classification in the Android platform. We specifically consider datasets exploited in the respective literature from 2012 onward. Then, we concentrate on the so far most commonly used and modern datasets, namely Drebin \cite{Drebin}, VirusShare \cite{VirusShare}, and AndroZoo \cite{AndroZoo}, and try to answer the second question per dataset. Precisely, by using the average coefficients of permissions and intents for a large number of malware instances per corpus, we demonstrate the most significant feature category. Lastly, we report the top ten features per dataset and discuss similarities between these corpora.

The remainder of this paper is organized in the following manner. The next section discusses the related work. Section \ref{datasets} details on the different datasets used to evaluate mobile malware detection approaches. Section \ref{featureImportance} provides our results on feature importance. The last section concludes and provides pointers to future work.

\section{Related work}
\label{relatedWork}

This section presents previous work on feature importance and feature selection. As already mentioned, thus far, this topic has received little attention in the literature.

Feizollah et al. \cite{Feizollah} categorized the available features into four groups, namely, static, dynamic, hybrid, and app's metadata. Furthermore, the authors evaluate the aforementioned features with regard to the difficulty of extraction and their popularity among the relevant literature. Finally, they offer a survey of the available datasets. On the downside, the only available datasets at the time of this research were Contagio \cite{Contagio}, MalGenome \cite{MalGenome}, and Drebin.

Zhao et al. \cite{Zhao} proposed a feature selection algorithm called \emph{FrequenSel}. According to the authors, FrequenSel selects features which are frequently used in malware and rarely used in benign apps, thus it can more accurately distinguish between the positive and negative class. During their experiments, the authors evaluated their approach with a collection of 7,972 apps, which contained malware collected from Drebin and other public malware libraries, as well as benign apps from Google Play. Their results reported an accuracy of up to 98\%. Similar to \cite{Feizollah}, the apps used in this work are nowadays considered outdated.

Kouliaridis et al. \cite{Kouliaridis} introduced an online open-source tool called \emph{Androtomist}, which performs hybrid analysis on Android apps. The authors focused on the importance of dynamic instrumentation, as well as the improvement in detection achieved when hybrid analysis is used vis-\`a-vis to static analysis. During their experiments, the authors compared feature importance between three datasets, namely Drebin, VirusShare, and AndroZoo. Finally, the authors elaborated on features which seem to be commonly exploited in malware and seldom in benign apps. While the datasets used in their work comprise newer apps as opposed to \cite{Feizollah} and \cite{Zhao}, the authors used a rather small subset of each dataset in the course of their experiments.

To the best of our knowledge, none of the above mentioned works address feature importance across multiple datasets with a large number of samples.

\section{Datasets}
\label{datasets}

Heretofore, several mobile app corpora have been built and exploited by researchers to evaluate malware detection approaches on the Android platform. This section surveys in chronological order all major mobile malware datasets used in the literature. Table \ref{datasetComparison} compares all datasets, with regard to their age, size, access, and impact to the research community. As shown in the table, AndroZoo and VirusShare are the only datasets still being updated today. The table also includes the number of works each dataset was employed according to: (i) a corresponding list of publications as given in the dataset's website, (ii) a listing of downloads in the dataset's website, and (iii) the citations the original dataset work, if any, has received according to Google Scholar.

\begin{itemize}

\item Contagio mobile mini-dump \cite{Contagio}: It is a publicly available repository of mobile malware samples. The samples were collected in 2010 and currently the dataset contains 189 malware samples, thus being by far the smallest available corpus.

\item MalGenome \cite{MalGenome}: In 2012, the MalGenome dataset was released. This corpus contains 1,260 malware samples categorized into 49 different malware families. The malware instances are dated from Aug. 2010 to Oct. 2011. The work which introduced this dataset seems to be by far the most highly cited. Unfortunately, the MalGenome project has stopped sharing their dataset in Dec. 2015. 

\item VirusShare \cite{VirusShare}: The access to the dataset's website is granted via invitation only. The dataset does not only contain mobile malware samples, but also samples from various platforms, including Windows and Linux. Furthermore, it is updated regularly and contains samples in the time span from 2012 to 2020. This dataset is also very popular in the research community, i.e., the number of works exploiting it is steadily growing every year.

\item Drebin \cite{Drebin}: It comprises 5,560 malware across 179 different families. The samples were collected between Aug. 2010 and Oct. 2012. Drebin is one of the most popular datasets and it is referenced in more that 1.3K works in the literature. On the downside, it has not received an update since 2012.

\item DroidBench \cite{DroidBench}: Is a set of apps implementing different types of data leakage. At present, the repository comprises 120 apps. The main task of these apps is data leak. Put simply, the samples in DroidBench are not real malware instances and are only meant to evaluate analysis tools.

\item PRAGuard \cite{PRAGuard}: It currently contains 10,479 malware samples, obtained by obfuscating the MalGenome and the Contagio mobile mini-dump datasets with seven different obfuscation techniques. The samples are dated from 2010 to 2011.

\item AndroZoo \cite{AndroZoo}: AndroZoo is a growing collection of Android apps collected from diverse sources, including the official Google Play store \cite{GooglePlay}. The dataset is updated regularly and it currently contains over 12M samples. The access to the dataset is granted by application only. The number of works using this dataset is also growing on a yearly basis.

\item Kharon \cite{Kharon}: It comprises only 7 instances of malware, namely, SimpLocker, BadNews, DroidKungFu1, SaveMe, MobiDash, WipeLocker, and Cajino, which have been manually dissected and documented. The samples are dated from 2012 to 2016.

\item Android Adware and General Malware Dataset (AAGM) \cite{AAGM}: It is generated from 1,900 apps belonging to the following three categories: 250 adware apps, 150 general malware apps, and 1,500 benign apps. Benign samples are dated from 2015 to 2016, but there is not enough information on the creation date of the malware samples.

\item AMD \cite{AMD}: It is a publicly shared dataset which contains 24,553 samples categorized in 135 varieties among 71 malware families. The samples are dated from 2010 to 2016. At the time of writing, the AMD website were unavailable.

\end{itemize}

\begin{table*}[h]
\centering
\footnotesize
\setlength{\tabcolsep}{3pt}
\def\arraystretch{1.4}
\begin{tabular}{ |l|c|c|r|l|r| } 
\hline
\textbf{Dataset} & \textbf{Created} & \textbf{Last updated} & \textbf{Size} & \textbf{Access type} & \textbf{Publications/Downloads/Citations} \\
\hline
Contagio mobile & 2010 & 2010 & 189 & Public & -/-/- \\
MalGenome & 2011 & 2011 & 1,260 & Unavailable & -/460/2181 \\
VirusShare & 2011 & 2020 & Unknown* & Invitation & 1307/-/- \\
Drebin & 2012 & 2012 & 5,560 & Public & -/157/1353 \\
DroidBench & 2013 & 2013 & 120 & Public & -/-/- \\
PRAGuard & 2015 & 2015 & 10,479 & Application & -/133/84 \\
AndroZoo & 2016 & 2020 & 12,498,250* & Application & -/-/267 \\
Kharon & 2016 & 2016 & 7 & public & -/-/20 \\
AAGM & 2017 & 2017 & 1,900* & Public & -/-/29 \\
AMD & 2017 & 2017 & 24,553 & Public & -/368/171 \\

\hline
\end{tabular}
\caption{Outline of major datasets ordered by their creation date. Asterisk = not all samples are malicious, Dash = Not available}
\label{datasetComparison} 
\end{table*}

\section{Feature importance}
\label{featureImportance}

A key factor that affects the accuracy of machine learning based malware detection methods is the importance of features contained in malware samples \cite{Kouliaridis}. To obtain a clear view of this aspect, the current section presents our results on feature importance over a great mass of malware apps collected from the state-of-the-art datasets. That is, as already pointed out in section \ref{datasets}, VirusShare and AndroZoo seem to be the only datasets still being updated today. Furthermore, the Drebin dataset has been used by a multitude of research works on the topic of mobile malware detection, thus making it ideal when comparing new detection methods with previous state-of-the-art.

Precisely, in the context of this section, we randomly collected 1K malware samples from each of these three datasets, as well as 1K random benign apps from Google play to create three 2K balanced datasets of both malware and benign apps. The samples are dated from 2010 to 2012, 2014 to 2017, and 2017 to 2020 for the Drebin, VirusShare and AndroZoo corpora, respectively. Static analysis was performed via the open-source tool \emph{Androtomist} \cite{Kouliaridis} to extract permissions and intents for each of the 3K malware plus 1K benign samples collected in total. Specifically, each app was decompiled to get the Manifest.xml file and log permissions and intents to create feature vectors, i.e., binary representations of each distinct feature.

The feature importance score is assigned by coefficients calculated as part of an Information Gain (IG) model. Specifically, IG is an entropy-based feature evaluation method and is defined as the amount of information provided by the feature items \cite{Lei}. Put simply, low probability, i.e., rare events are more surprising and have a greater amount of information. This also means that probability distributions where the events are almost equally likely are more surprising and have larger entropy. Therefore, in our case, information entropy can be roughly thought of as how much variance the data have. For example, a dataset of only one feature would have zero entropy. On the other hand, a dataset of mixed features would have relatively high entropy. The formula to calculate the information Entropy for a dataset with $C$ classes is as follows:
$$E = -\sum_{i}^{C} p_i log_2 p_i$$
where $p$ is the probability of randomly picking an element of class $i$, i.e., the proportion of the dataset made up of class $i$. It is worth noting that the entropy metric of uncertainty introduced by Shannon \cite{Shannon} has been exploited in several works in the information security literature \cite{Ehrlich2008, Tsiatsikas, Tsiatsikas2}.

Computing the IG for a feature involves calculating the entropy of the class label, i.e., positive (malware) or negative (benign) for the entire dataset and subtracting the conditional entropies for each possible value of that feature, in our case ``exist'' (1) or ``not exist'' (0). The entropy calculation requires a frequency count of the class label by feature value. Precisely, the instances of a dataset are selected with a feature value $x$. Then, the occurrences of each class are counted and the entropy for $x$ is computed. This step is repeated for each possible value $x$ (0,1) of the feature. The formula to calculate IG is as follows:
$$IG(D, v) = H(D) – H(D | v)$$
Where $IG(D, v)$ is the information gain for the dataset $D$ for the variable $v$, $H(D)$ is the entropy for the dataset before any change, and $H(D | v)$ is the conditional entropy for the dataset given the variable $v$. The higher the IG score the more information is gained from this feature.

Tables \ref{D_FI}, \ref{VS_FI}, and the left side of table \ref{AZ_FI} include the top 10 features observed for Drebin, VirusShare, and AndroZoo, respectively, along with their IG score. By observing the top 10 features of Drebin and VirusShare in Tables \ref{D_FI} and \ref{VS_FI}, it becomes obvious that there is a similarity between the top features of these corpora. More specifically, the first three features are the same for both Drebin and VirusShare's top 10. In total, 7 out of 10 features are identical in both tables divided into 6 permissions and 1 intent. On the other hand, as shown in the left side of table \ref{AZ_FI}, AndroZoo has 1 out of 10 identical features with Drebin's top 10, and shares zero out of 10 identical features with VirusShare's top 10. Lastly, all of the Androzoo's top 10 features are intents, contrariwise to Drebin and VirusShare where only 2 and 3 out of 10 are intents, respectively. This further demonstrates the difference in feature importance among the examined datasets.

To verify our conclusions on feature importance regarding the examined datasets, we randomly selected an additional 1K malware apps from the most contemporary one, i.e., Androzoo, and also randomly added 1K new benign apps from Google Play. This doubles the number of instances contained in our AndroZoo dataset, i.e., 2K malware and 2K benign apps in total. The right side of table \ref{AZ_FI} contains the feature importance scores for this new double size dataset. As expected, the top 10 features in Table \ref{AZ_FI} are all intents too. Also, 8 out of 10 features are common to the two sides of the table, and 3 out of the 4 top features occupy the same places in both sides of the table. This further supports the observation that feature importance is tightly related to the age of the malware. Naturally, this phenomenon may negatively affect the performance of older detection methods if solely based on these two categories of features.

\begin{figure*}[h]
  \centering
  \includegraphics[width=0.75\textwidth]{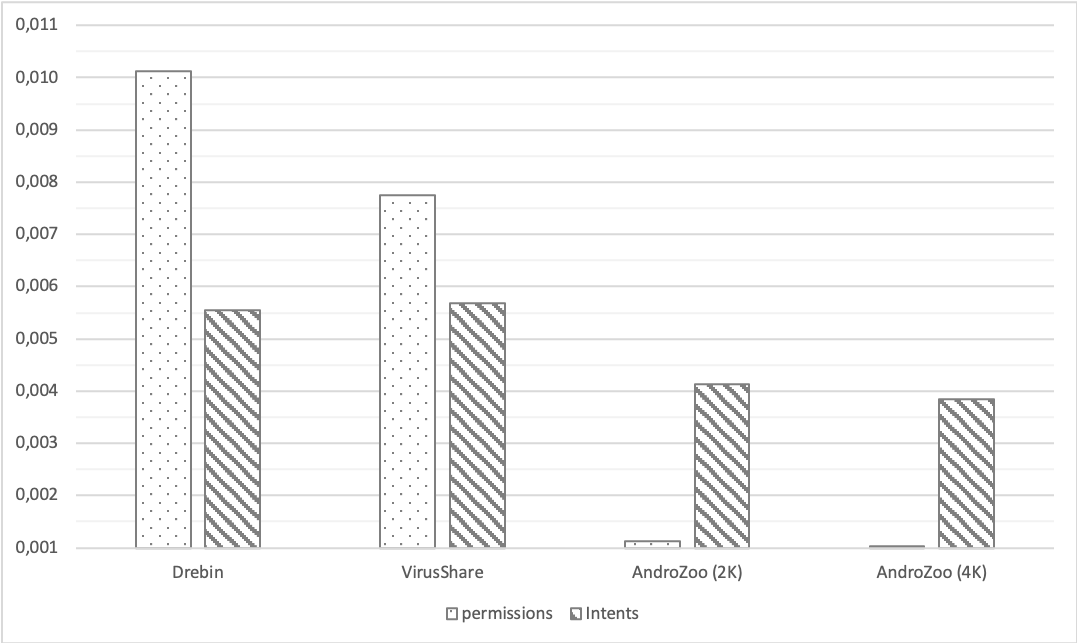}
  \caption{Average feature importance scores on all three datasets for the two feature categories (Permissions and Intents).}
  \label{comparison}
\end{figure*}

Figure \ref{comparison} illustrates the average feature importance scores per dataset, for both the examined feature categories, namely permissions and intents. Note that the mean score is calculated over all the permissions and intents identified, and not solely on the top 10 values included in tables \ref{D_FI}, \ref{VS_FI}, and \ref{AZ_FI}. As easily observed from the figure, in the AndroZoo corpus, intents produced a much more higher - approximately triple - IG score than permissions. Emphatically, this situation applies almost equally to both the 2K and 4K datasets. Nevertheless, this picture is clearly inverted in the Drebin and VirusShare corpora, that is, in Drebin there is an $\approx$ +0.0045 and in VirusShare an $\approx$ +0.002 higher score than that of intents. 

\begin{table}[H]
\centering
\scriptsize
\setlength{\tabcolsep}{3pt}
\def\arraystretch{1.5}
\begin{tabular}{ |cll| } 
\hline
\textbf{IG Score} & \textbf{Feature} & \textbf{Category} \\ 
\hline
0.2294 & android.permission.INTERNET & Permissions \\ 
0.2130 & android.permission.READ\_PHONE\_STATE	& Permissions \\ 
0.1335 & android.permission.SEND\_SMS & Permissions \\ 
0.0994 & android.permission.WRITE\_EXTERNAL\_STORAGE & Permissions \\ 
0.0965 & android.permission.RECEIVE\_BOOT\_COMPLETED & Permissions \\ 
0.0939 & android.permission.RECEIVE\_SMS & Permissions \\ 
0.0857 & android.permission.READ\_SMS & Permissions \\ 
0.0810 & android.intent.action.BOOT\_COMPLETED	& Intents \\ 
0.0706 & com.google.android.c1dm.intent.RECEIVE & Intents \\ 
0.0683 & android.permission.ACCESS\_COARSE\_LOCATION & Permissions \\
\hline
\end{tabular}
\caption{Top 10 features in the Drebin dataset.}
\label{D_FI} 
\end{table}

\begin{table}[H]
\centering
\scriptsize
\setlength{\tabcolsep}{3pt}
\def\arraystretch{1.5}
\begin{tabular}{ |cll| } 
\hline
\textbf{IG Score} & \textbf{Feature} & \textbf{Category} \\ 
\hline
0.2305 & android.permission.INTERNET & Permissions \\ 
0.2276 & android.permission.READ\_PHONE\_STATE	& Permissions \\ 
0.1713 & android.permission.SEND\_SMS & Permissions \\ 
0.1477 & android.permission.RECEIVE\_SMS & Permissions \\ 
0.1328 & android.permission.WRITE\_EXTERNAL\_STORAGE & Permissions \\ 
0.1067 & android.permission.READ\_SMS & Permissions \\ 
0.0958 & android.intent.category.HOME & Intents \\ 
0.0926 & android.intent.action.DATA\_SMS\_RECEIVED	& Intents \\ 
0.0648 & android.intent.action.BOOT\_COMPLETED	& Intents \\ 
0.0610 & android.permission.WAKE\_LOCK	& Permissions \\
\hline
\end{tabular}
\caption{Top 10 features in the VirusShare dataset.}
\label{VS_FI} 
\end{table}

\begin{table*}[ht]
\centering
\tiny
\setlength{\tabcolsep}{3pt}
\def\arraystretch{1.5}
\begin{tabular}{ |cl||cl| } 
\hline
\textbf{IG Score} & \textbf{Feature} & \textbf{IG Score} & \textbf{Feature}  \\ 
\hline
0,1550 & android.intent.action.USER\_PRESENT & 0,1680	& android.intent.action.USER\_PRESENT  \\

0,1401 & android.intent.action.PACKAGE\_REMOVED & 0,1528	& android.intent.action.PACKAGE\_REMOVED  \\

0,1208 & android.intent.category.DEFAULT & 0,1208	& android.intent.category.BROWSABLE  \\

0,0769 & android.intent.action.PACKAGE\_ADDED & 0,1162	& android.intent.action.PACKAGE\_ADDED  \\

0,0672 & android.intent.category.BROWSABLE & 0,0955	& cn.jpush.android.intent.NOTIFICATION\_RECEIVED\_PROXY  \\

0,0652 & android.intent.action.VIEW & 0,0812 & android.intent.action.ACTION\_POWER\_CONNECTED  \\

0,0582 & com.google.android.c1dm.intent.RECEIVE & 0,0780	& org.agoo.android.intent.action.RECEIVE  \\

0,0530 & cn.jpush.android.intent.NOTIFICATION\_RECEIVED\_PROXY & 0,0722	& com.google.android.c1dm.intent.RECEIVE  \\

0,0521 & android.intent.action.ACTION\_POWER\_CONNECTED & 0,0685	& android.intent.action.MEDIA\_MOUNTED  \\

0,0518 & org.agoo.android.intent.action.RECEIVE & 0,0685	& cn.jpush.android.intent.NOTIFICATION\_OPENED  \\
\hline
\end{tabular}
\caption{Top 10 features in the AndroZoo dataset, all features are intents. Left: 2K apps dataset, right: 4k apps dataset}
\label{AZ_FI} 
\end{table*}

\section{Conclusions}
\label{conclusions}

This work examined the literature on Android malware detection spanning the period from 2012 to 2020. The focus was on contributions exploiting machine learning and on identifying the datasets used in each relevant work. Our analysis showed that three datasets, namely Drebin, VirusShare, and AndroZoo stand out. Following, we used a significant mass of malware instances existing in each of the aforementioned datasets along with a large number of benign instances to estimate the feature importance of permissions and intents. We reported the most important features per dataset in terms of IG, as well as similarities and differences between the top features in each of them. Our results reveal a noteworthy difference in feature importance when inspecting our partial AndroZoo datasets vis-\`a-vis the other two. As a future work, we aim to improve this research by also examining the significance of features stemming from dynamic analysis. Moreover, based on feature importance, we aim to examine feature dimension reduction via the use of diverse techniques.

\bibliography{main.bib}

\end{document}